\documentclass[twocolumn,floatfix,groupedaddress,superscriptaddress,amsmath,a4paper,twoside,showkeys]{revtex4}
\pdfoutput=1
\usepackage{graphicx}
\usepackage{chemarr}
\usepackage{times,longtable}
\usepackage{hyperref,amsmath}
\usepackage{version}
\usepackage{color}
\usepackage{soul,amssymb}

\usepackage{epsf,mathtools}
\usepackage[usenames,dvipsnames]{xcolor}

\graphicspath{{figures/},{figures/pdf/},{figures/eps/}}

\bibpunct{[}{]}{,}{n}{}{,}

\providecommand{\addhyphen}[1]{#1.---}

\makeatletter
\renewcommand \paragraph{%
  \@startsection
    {paragraph}%
    {4}%
    {\parindent}%
    {\z@}%
    {-.1em}%
    {\normalfont\normalsize\itshape\addhyphen}%
}%
\makeatother

\begin{document}

\title{Identifying all irreducible conserved metabolite pools in genome-scale metabolic networks: a general method and the case of Escherichia coli}
 

\author{A. De Martino}

\altaffiliation{All authors contributed equally.}

\affiliation{IPCF-CNR, Unit\`a di Roma-Sapienza, Roma (Italy)}

\affiliation{Dipartimento di Fisica, Sapienza Universit\`a di Roma, p.le A. Moro 2, 00185 Roma (Italy)}

\affiliation{Center for Life Nano Science@Sapienza, Istituto Italiano di Tecnologia, Viale Regina Elena 291, 00161 Roma (Italy)}

\author{D. De Martino}

\altaffiliation{All authors contributed equally.}

\affiliation{Center for Life Nano Science@Sapienza, Istituto Italiano di Tecnologia, Viale Regina Elena 291, 00161 Roma (Italy)}

\author{R. Mulet}

\altaffiliation{All authors contributed equally.}

\affiliation{Henri-Poincar\'e-Group of Complex Systems and Department of Theoretical Physics, Physics Faculty, University of Havana, CP 10400 La Habana (Cuba)}

\author{A. Pagnani}

\altaffiliation{All authors contributed equally.}

\affiliation{DISAT and Centre for Computational Sciences, Politecnico di Torino, Corso Duca degli Abruzzi 24, 10129 Torino (Italy)}

\affiliation{Human Genetics Foundation, Via Nizza 52, 10126 Torino (Italy)}

\begin{abstract}

The stoichiometry of metabolic networks usually gives rise to a family of conservation laws for the aggregate concentration of specific pools of metabolites, which not only constrain the dynamics of the network, but also provide key insight into a cell's production capabilities. When the conserved quantity identifies with a chemical moiety, extracting all such conservation laws from the stoichiometry amounts to finding all integer solutions to an NP-hard programming problem. Here we propose a novel and efficient computational strategy that combines Monte Carlo, message passing, and relaxation algorithms to compute the complete set of irreducible integer conservation laws of a given stoichiometric matrix, also providing a {\em certificate} for correctness and maximality of the solution. The method is deployed for the analysis of the complete set of irreducible integer pools of two large-scale reconstructions of the metabolism of the bacterium {\it Escherichia coli} in different growth media. In addition, we uncover a scaling relation that links the size of the irreducible pool basis to the number of metabolites, for which we present an analytical explanation. 
\end{abstract}



\maketitle

\section{Introduction}
\label{sec:intr}

When studying metabolic networks at the scale of the whole genome, it is often the case that the information required to develop dynamical models is not available, because either kinetic parameters or reaction mechanisms are unknown. In most cases, the only reliable information is encoded in the stoichiometry of the reaction network \cite{Jeong,Fell,Zhu,Palssonnet}. Over the past decade or so, it has gradually become clear that the stoichiometric matrix by itself harbors a host of important physical, biological and functional information that can be extracted by combining different algebraic and computational tools \cite{palssonbook,beardbook}. Constraint-based models like Flux Balance Analysis (FBA), for instance, have shed light on functional optimality in different contexts, providing, in many cases (mostly for unicellular organisms) unprecedented predictive power  \cite{palssonbook}. Other examples of stoichiometry-based features that have been proposed and investigated include extreme pathways \cite{Yeung,Papin}, flux modes \cite{Schuster}, the geometry of the space of flux configurations compatible with a non-equilibrium steady state assumption for metabolite levels \cite{MC1,MC2,Braunstein}, and the corresponding distributions of allowed fluxes \cite{Braunstein,Carlotta,MC3}. At odds with FBA, which can now almost routinely be performed on networks of any size \cite{COBRA},  tackling the issues listed above on genome-scale networks with thousands of reactions and metabolites presents serious computational challenges, as the algorithms currently available do not scale gently with the system size. This is also the case for the problem of identifying the so-called conserved metabolite pools that we shall consider here \cite{Schuster91,Famili}. 

In brief, conserved pools are particular types of conservation laws that can be identified in biochemical networks. It is a well known algebraic fact that, as a result of the sheer structure of the input-output stoichiometry, given a stoichiometric matrix one can find linear combinations of concentration variables that are due to be constants of motion of the dynamical system governing the time evolution of concentrations and reaction rates. The existence of such laws has profound consequences. In first place, any intervention aimed at altering the level of a certain metabolite should consider whether its variations are limited or not by conservation relationships. Secondly, such conservation laws constrain a network's production capabilities, as there clearly cannot be a net production of a compound belonging to a conserved quantity. Therefore, as long as one does not consider {\it ad hoc} sink reactions like biomass production, mapping out these conservation laws amounts to obtaining a genome-scale picture of what a cell can (in principle) excrete or make available to processes outside metabolism, such as protein synthesis. Finally, conservation laws are crucial for the dynamics of metabolic networks, since they imply an effective reduction of the number of independent flux or concentration variables.

The problem of finding conservation laws is relatively straightforward to solve with the tools of linear algebra, since, as said above, conservation laws correspond to specific (linear) dependencies of the rows of the stoichiometry matrix. There are important cases, however, when the issue takes a more challenging twist, namely when one is interested in the conservation of specific chemical moieties. It is simple to understand that, because of the intrinsic discreteness of moieties, the combinations that describe their conservation should only be constructed with non-negative integers. These specific conservation laws correspond to what we shall call here `conserved metabolite pools' (CMPs). CMPs have been shown to have relevant biotechnological or medical implications: for example, a few years ago, Bakker et al. \cite{Bakker} revealed the role played by moiety conservation in the protection of trypanosomes. In passing from a linear to an integer programming problem, however, the level of complexity increases (in our case, finding CMPs is  known to fall within the NP-hard class \cite{papadibook,pasqualina}). Yet more difficult is the identification of {\it all} irreducible CMPs, i.e. of all pools that cannot be expressed as combinations of other pools, which is equivalent to finding all irreducible solutions of an integer programming problem. 

It is perhaps not surprising, then, that progress in this direction has been relatively slow \cite{Rao,Sauro}. In 2003 Famili and Palsson \cite{Famili} have shown that a convex representation of the left null space of the stoichiommetric matrix determines the metabolic pools. However, their proposed computational approach scales exponentially with the system size, and can only be employed for the analysis of rather small networks. In 2005 Nikolaev et al. \cite{Nikolaev} introduced the Metabolite Concentration Coupling Analysis (MCCA) and the Minimal Pool Identification (MPI) tools for genome-scale metabolic networks. MCCA allows for the identification of subsets of metabolites whose concentrations are coupled within common conserved pools, while MPI helps to determine the conserved pools for individual metabolites. Imielinski and coworkers \cite{Imielinski} have instead exploited the formal algebraic duality of metabolite producibility and conservation to devise a method that relates biomass producibility to nutrient availability, which they then applied to the metabolism of {\it Escherichia coli}, obtaining a large set of novel putative growth media. Despite such efforts, a consistent computational tool to determine {\em all} irreducible conserved pools in a genomic scale networks has remained so far elusive. 

In this work we construct such a computational method. The technique we propose exploits the above mentioned duality and combines different kinds of algorithms (message passing, Monte Carlo and relaxation). Its background and structure are discussed in Sec. II. As case studies, we have considered different reconstructions of the metabolic network of the bacterium {\it E. coli}. In particular, we have been able to identify in each case {\it all} irreducible CMPs in different conditions (Sec. III). By studying {\it E. coli}, moreover, we have uncovered a relation between the number of irreducible pools and the size of a network (number of metabolites and/or reactions), a theoretical justification for which is also discussed in Sec. III. Finally, details about the algorithms are presented in Sec. IV, while our conclusions and perspectives are reported in Sec. V.

\section{The problem and the solution strategy}
\label{sec:meth}


Given a metabolic network encoded by the stochiometric matrix $\mathbb{S} = (S_{mr})$, where $S_{mr}$ is the stochiometric coefficient of metabolite $m \in\{ 1, \dots, M\}$ in reaction $r\in\{1, \dots, N\}$ (with the standard sign convention to distinguish substrates from products), the time evolution of the concentration vector $\mathbf{c}=\{c_m\}$ satifies
\begin{equation}\label{dyn}
\dot{\mathbf{c}}=\mathbb{S}\mathbf{v}~~,
\end{equation}
where $\mathbf{v}=\{v_r\}$ is the vector of reaction fluxes and we have assumed that the stoichiometry of metabolite exchanges with the environment is included in $\mathbb{S}$. In non-equilibrium steady states with clamped concentrations, fluxes arrange in such a way that $\mathbb{S}\mathbf{v}=\mathbf{0}$, and solutions with $\mathbf{v\neq 0}$ correspond to non-equilibrium steady flux states. Now consider a linear combination of
concentration variables with fixed coefficients $k_m\geq 0$, i.e.
\begin{equation}
\ell=\sum_{m=1}^M k_m c_m\equiv (\mathbf{k},\mathbf{c})~~~~~,~~~~~
\mathbf{k}=\{k_1,\dots,k_M\}~~.
\end{equation}
Clearly,
\begin{equation}
\dot{\ell}=(\mathbf{k},\dot{\mathbf{c}})=(\mathbf{k},\mathbb{S}\mathbf{v})=(\mathbf{v},\mathbb{S}^T\mathbf{k})~~.
\end{equation}
If $\mathbf{k}$ belongs to the left null-space of $\mathbb{S}$, that is if 
\begin{eqnarray}
\label{sys1}
& &\mathbb{S}^T\mathbf{k}=\mathbf{0}\\
& &  \mathbf{k} \neq {\bf 0}   \,\,\, , \,\,\,  k_m \geq 0 \,\,\, \forall
m\in \{1,\dots,M\} 
\label{sysg}
\end{eqnarray}
then the aggregate concentration variable $\ell$ is conserved in {\it any} flux state $\mathbf{v}$. We shall generically call a combination defined by a vector $\mathbf{k}$ satisfying (\ref{sys1})  a ``conservation law''. From a physical viewpoint, conservation laws represent constraints  for aggregate levels 
that are required to be satisfied by trajectories of the dynamics of the system, i.e. by (\ref{dyn}) with given specifications of how $\mathbf{v}$ depends on $\mathbf{c}$. According to the above definition,
\begin{equation}\label{bound}
\#\{\textrm{independent laws}\} \leq M-\textrm{rank}(\mathbb{S}) ~~.
\end{equation} 
Note that, in principle, every vector $\mathbf{k}$ belonging to the
left kernel of $\mathbb{S}$ and not limited by the requirements
(\ref{sysg}) defines a conservation law, and the total number of
linearly independent laws of such type equals the dimension of the
null-space of $\mathbb{S}$, i.e. $M-\textrm{rank}(\mathbb{S})$. The restriction to $\mathbf{k\geq 0}$ in (\ref{sysg}), which causes the inequality in (\ref{bound}),
allows for a more straightforward physical interpretation of
quantities like $\ell$. Consider, for instance, the toy network formed by the three reactions 
\begin{gather*}
A+B\to C+D\\
E\to F+B \\ 
D+F\to E
\end{gather*}
One easily checks that it possesses two conservation laws, i.e. $\ell_1= c_B+c_E+c_D$ and $\ell_2= c_E+c_F$. Clearly, $\ell_2$ could be also written as $\ell'_2= c_F-c_B-c_D$. While both $\ell_2$ and $\ell'_2$ describe conservation laws, $\ell_2$ can be interpreted as the conservation of a total enzyme mass if $E$ and $F$ are seen to represent, respectively, a bound and a free enzyme species. A similar physical interpretation is harder to find for $\ell'_2$ (and the situation rapidly becomes more complicated in larger networks). In particular, among solutions of (\ref{sys1}), those for which $k_m$ are {\em non-negative integers} can be fully rationalized in chemical terms as related to the conservation of moieties, groups or chemical elements. In what follows we shall focus on these, and define hereafter a CMP as a solution of
\begin{eqnarray}
\label{sys2}
& &\mathbb{S}^T\mathbf{k}=\mathbf{0}\\
& & \mathbf{k} \neq {\bf 0}  \quad, \quad k_m\in\{0,1,2,\ldots\}~\forall m~~.\label{sysf}
\end{eqnarray}
When such $\mathbf{k}$'s suffice to generate all conservation laws, then the number of independent CMPs saturates the bound (\ref{bound}). 

The problem we face here concerns the identification of all {\it irreducible} CMPs of a given stoichiometric matrix $\mathbb{S}$, where a CMP is said to be irreducible if it cannot be expressed as a sum of other pools with positive coefficients. Knowing irreducible pools then amounts to disposing of a basis through which all possible conservation laws of the network can be characterized, provided the bound (\ref{bound}) is saturated (which, as we shall see, is not always the case due to the positivity restriction in (\ref{sysg})). This problem is unluckily NP-hard \cite{Coleman_Pothen} and deterministic algorithms fail when the underlying network is sufficiently large. In particular, for the sizes relevant to genome-scale metabolic modeling ($N,M\gtrsim 10^3$) one always runs into the combinatorial explosion of computation times. Stochastic strategies are therefore mandatory. In brief, we shall map this task to a global optimization problem whose solution can be retrieved via stochastic algorithms known to be exact in special situations, a kind of approach that has been used before with considerable success in the solution of other NP problems \cite{BMZ,ZecchinaBraunstein,Coloring,Braunstein}.

Our strategy is divided in three steps: (a) compute a list of all metabolites belonging to at least one pool; (b) extract individual pools from that list; (c) check that $\mathbb{S}$ does not allow for any further pools. Step (a) will be carried out by a message-passing procedure, step (b) by Monte Carlo, and step (c) by a relaxation algorithm. The outline is as follows (see `Materials and methods' for more details). As in \cite{Imielinski}, we shall exploit the connection between (\ref{sys2}) and its dual system \cite{Schuster91}, i.e.
\begin{eqnarray}\label{sys3}
& & \mathbb{S}\mathbf{v}\geq 0\\
& & \mathbf{v} \neq \mathbf{0} \qquad,\qquad v_r\in\mathbb{R}~~\forall r ~~.
\end{eqnarray}  
The solution spaces of (\ref{sys2}) and (\ref{sys3}) are connected by the Motzkin theorem of the alternative, which can be stated as follows:

{\bf Theorem} (Motzkin, 1936). {\it Consider any arbitrary subset $R$ of rows of $\mathbb{S}$. Then, either there exists a solution $\mathbf{v^*}$ to system (\ref{sys3}) such that all inequalities corresponding to the subset $R$ hold strictly, or system (\ref{sys2}) has a solution $\mathbf{k^*}$, with $k_m^*>0$ for each $m \in R$.}

In essence, Motzkin's result guarantees that solutions of (\ref{sys3}) verify strict {\it equalities} for metabolites belonging to CMPs. This is rather intuitive if one interprets strict inequalities in (\ref{sys3}) as conditions for metabolite producibility \cite{Imielinsk05,Imielinski}.  Luckily, a solution of a subset of constraints in (\ref{sys3}) with strict {\it inequalities} can be found very efficiently by relaxation algorithms (e.g. MinOver \cite{jstat} or Motzkin method \cite{shrij}). Therefore a simple numerical check to confirm that all independent pools have been found
consists in looking for a solution of (\ref{sys3}) with strict inequalities for all $m$'s remaining after having removed from $\mathbb{S}$ the rows corresponding to metabolites belonging to at least one CMP. If no solution is found, then the reduced $\mathbb{S}$ necessarily harbors more pools. 

Disposing of a polynomial algorithm that confirms whether all the elements belonging to at least one pool have been found, what one needs is a fast strategy to find them. To extract a list of conserved metabolites we shall resort to Belief Propagation (BP), a message-passing technique that is exact on trees and has already proved helpful in the study of metabolic networks \cite{Braunstein} (see `Materials and methods' for details). If the list is complete, relaxation will (as said above) converge to a solution of (\ref{sys3}) for the reduced stoichiometric matrix. Otherwise, the list derived from BP is incomplete and one should look for additional conserved metabolites in the reduced $\mathbb{S}$. This is easily done by analyzing the dynamics of the relaxation algorithm, as previously done for the mathematically related problem of identifying thermodynamically infeasible cycles in flux patterns \cite{noiplos}. Once a complete list is available, one may disentangle irreducible CMPs by defining a cost function with minima corresponding to the solutions of (\ref{sys2},\ref{sysf}) and running a Monte Carlo minimization to look for all solutions (see `Materials and methods' for details). This last step will be computationally affordable thanks to the fact that the number of variables (equal to the number of metabolites in the final list) will be much smaller than $M$.

 \section{Results}
\label{sec:resu}

We shall analyze here the CMPs found for two reconstructions of {\it E. coli}'s metabolic networks of rather different sizes, namely iJR904 \cite{Palssonnet}, with $M=761$ and $N=1074$, and iAF1260 \cite{Feist}, with $M=1668$ and $N=2381$. These numbers refer to the sizes of the respective stoichiometric matrices with all uptakes and without the biomass reaction. Two limiting cases for the choice of the exchange fluxes will be considered. First, we shall analyze CMPs formed in a `rich medium', where all uptake reactions are active. Then, we shall look at the case of `minimal medium', by studying CMPs after having eliminated part of the uptakes. (In the latter case a much larger number of CMPs is to be expected.) We shall see that, while for iAF1260 irreducible CMPs suffice to generate all independent conservation laws, i.e. the bound (\ref{bound}) is saturated, the model iJR904 presents one conservation law that cannot be derived from CMPs (which however violates (\ref{sysg})).


\subsection{iAF1260}

In Table \ref{pool} we show the CMPs found for iAF1260 with a `rich medium'. They are 38 in total, matching exactly the dimension of the left kernel of the stoichiometric matrix. 

\begin{table*}
\begin{tabular}{  | c  | c  |  l | p{12.5cm} | }  
\hline 
CMP ID & Size & Conserved species & Formula  \\ 
\hline 
1--19 & 2 & tRNA &  alatrna[c]  $+$  trnaala[c], 
 argtrna[c]  $+$  trnaarg[c],
 asntrna[c]  $+$  trnaasn[c],
 asptrna[c]  $+$  trnaasp[c],
 cystrna[c]  $+$  trnacys[c],
 glntrna[c]  $+$  trnagln[c],
 glutrna[c]  $+$  trnaglu[c],
 glytrna[c]  $+$  trnagly[c], 
 histrna[c]  $+$  trnahis[c],
 iletrna[c]  $+$  trnaile[c],
 leutrna[c]  $+$  trnaleu[c],
 lystrna[c]  $+$  trnalys[c],
 phetrna[c]  $+$  trnaphe[c],
 protrna[c]  $+$  trnapro[c],
 sertrna[c]  $+$  trnaser[c],
 thrtrna[c]  $+$  trnathr[c],
 trptrnatrp[c]  $+$  trnatrp[c],
 tyrtrnatyr[c]  $+$  trnatyr[c],
 valtrnaval[c]  $+$  trnaval[c] \\
20--23 & 2 &  missing transport and leaves &   
arbt6p[c]  $+$  hqn[c],
cyan[c]  $+$  tcynt[c], 
dms[c]  $+$  dmso[c],
tma[c]  $+$  tmao[c]  \\
24--28 & 2 &  lipoprotein & alpp[p]  $+$  lpp[p], 
dsbaox[p]  $+$  dsbard[p],  
dsbcox[p]  $+$  dsbcrd[p],
dsbdox[c]  $+$  dsbdrd[c],
dsbgox[p]  $+$  dsbgrd[p]  \\
29--31 & 2 &  redox enzymes &  fldox[c]  $+$  fldrd[c],
grdox[c]  $+$  grxrd[c],
trdox[c]  $+$  trdrd[c]   \\  
32 & 3 &  tRNA  &  fmettrna[c]  $+$  mettrna[c]  $+$  trnamet[c]    \\  
33--34 & 3 &  selenium compounds &  sectrna[c]  $+$  seln[c]  $+$  selnp[c],
 sectrna[c]  $+$  sertrnasec[c]  $+$  trnasecys[c]  \\
35 & 3 &  biotin &  btn[c]  $+$  btnso[c]  $+$  s[c]   \\
36 & 3 &  ~ &  8aonn[c]  $+$  amob[c]  $+$  pmcoa[c]    \\  
37 & 6 &  ~ &  8aonn[c]  $+$ btn[c]  $+$  btnso[c]  $+$  dann[c]  $+$  dtbt[c]  $+$  pmcoa[c]   \\  
38 & 53 &  ACP &  3haACP[c]  $+$  3hcddec5eACP[c]  $+$  3hcmrs7eACP[c]  $+$  3hcpalm9eACP[c]  $+$  3hcvac11eACP[c]  $+$  3hddecACP[c] $+$ 
              3hdecACP[c]  $+$  3hhexACP[c]  $+$  3hmrsACP[c]  $+$  3hoctACP[c]  $+$  3hoctaACP[c]  $+$  3hpalmACP[c]  $+$  
            3ocddec5eACP[c]  $+$  3ocmrs7eACP[c]  $+$  3ocpalm9eACP[c]  $+$  3ocvac11eACP[c]  $+$  3oddecACP[c]  $+$  3odecACP[c] $+$   
              3ohexACP[c]  $+$  3omrsACP[c]  $+$  3ooctACP[c]  $+$  3ooctdACP[c]  $+$  3opalmACP[c]  $+$  ACP[c]  $+$  acACP[c]  $+$  
            actACP[c]  $+$  apoACP[c]  $+$  but2eACP[c]  $+$  butACP[c]  $+$  cddec5eACP[c]  $+$  cdec3eACP[c]  $+$  dcaACP[c]  $+$  
            ddcaACP[c]  $+$  hdeACP[c]  $+$  hexACP[c]  $+$  malACP[c]  $+$  myrsACP[c]  $+$  ocACP[c]  $+$  ocdcaACP[c]  $+$  octeACP[c]  $+$    
            palmACP[c]  $+$  t3c11vaceACP[c]  $+$  t3c5ddeceACP[c]  $+$  t3c7mrseACP[c]  $+$  t3c9palmeACP[c]  $+$  tddec2eACP[c]  $+$  tdeACP[c]  $+$ 
            tdec2eACP[c]  $+$  thex2eACP[c]  $+$  tmrs2eACP[c]  $+$  toct2eACP[c]  $+$  toctd2eACP[c]  $+$  tpalm2eACP[c] \\
\hline
\end{tabular}
\caption{The $38$ CMPs found for the network iAF1260 in a `rich medium'. The suffixes [c] and [p] indicate the presence of that species in the cytoplasm and periplasm, respectively, in agreement with the compartmentation indicated in the reconstruction.}
\label{pool}
\end{table*}

$20$ of them (numbers 1--19 and 32) are formed by a tRNA in two forms: free and bound to its corresponding amino-acid. To have a physical interpretation, we note that if a model possesses a CMP corresponding to a moiety conservation, then that model is closed with respect to that moiety, in the sense that it does not allow for changes in the level of that particular chemical group. In this sense, CMPs based on a tRNA reflect the fact that, in the model where they have been found, the expression of each tRNA is necessarily constant (more precisely, it is assumed to change on time scales longer than those over which metabolite levels equilibrate). 

Compounds in CMP 20, arbutin 6-phosphate (arbt6p) and hydroquinone (hqn) are `leaves' of this network, since they only occur in one reaction (arbutin 6-phosphate glucohydrolase: arbt6p $+$ h2o $\to$ g6p $+$ hqn). 
CMP 21 is composed by hydrogen cyanide (cyan) and thiocyanate (tcynt) in their citoplasmic form. Interestingly these compounds are conserved despite the fact that, in the rich medium, there are uptakes for both. This is because the model lacks reactions that transport the periplasmic species into the cytoplasm. Exactly the same situation holds for the CMPs 22 and 23 formed by dymethil-sulfide (dms) and -sulfoxide (dmso), and by thrymethilammine (tma) and thrymethilammine-N-oxide (tmao). 

CMPs 24--28 express the conservation of the lipoproteins (apolipoprotein, disulfide isomerase I and II, disulfide interchange and oxidase). Notice that CMP 28 is the only one involving periplasmic species only. 

CMPs 29--31 describe the conservation of the redox enzymes flavodoxin (fldox), glutaredoxin (grdox) and thioredoxin (trdox). 

The compounds in pools 33 and 34 are all based on the element selenium, with respect to which the model is closed (i.e. there are no uptakes of compounds containing selenium). We also note from this example that CMPs can be overlapping, as one compound (sectrna) belongs to two different pools.


CMP 35 reflects the conservation of biotin, while the sulfur atom is a leaf of the network, appearing only in the  biotin synthase reaction. 

CMP 36 express the conservation law among 8-amino-7-oxononanoate (8aonn), S-adenosyl-4-methylthio-2-oxobutanoate (amob) and pimeloyl-coa (pmcoa), while CMP 37 involves 8aonn, pmcoa, the biotin compounds of pool 35 (btn, btnso), plus 7,8-diaminononanoate (dann) and dethiobiotin (dtbt), providing a further instance of overlapping irreducible pools. 

Finally, CMP 38 represents the conservation of the acyl carrying protein (ACP).

\begin{table*}
\begin{tabular}{  | c  | c  |  p{15cm} | }  
\hline 
Pool ID & Size & Formula  \\ 
\hline 
39  &  2  &  ag[c] $+$  ag[e] \\ 
40--47  &  3  &  cd2[c] $+$  cd2[e] $+$  cd2[p], 
  ni2[c] $+$  ni2[e] $+$  ni2[p],
  mobd[c] $+$  mobd[e] $+$  mobd[p], 
  cobalt2[c] $+$  cobalt2[e] $+$  cobalt2[p], 
 tungs[c] $+$  tungs[e] $+$  tungs[p], 
 met-D[c] $+$  met-D[e] $+$  met-D[p],
 hg2[c] $+$  hg2[e] $+$  hg2[p],
 cl[c] $+$  cl[e] $+$  cl[p]\\
48--57  &  4  &  betald[c] $+$  glyb[c] $+$  glyb[e] $+$  glyb[p],
bbtcoa[c] $+$  gbbtn[c] $+$  gbbtn[e] $+$  gbbtn[p], 
4hoxpacd[e] $+$  4hoxpacd[p] $+$  tym[e] $+$  tym[p],  
dms[e] $+$  dms[p] $+$  dmso[e] $+$  dmso[p],
cyan[e] $+$  cyan[p] $+$  so3[e] $+$  so3[p],
3sala[c] $+$  so2[c] $+$  so2[e] $+$  so2[p], 
gdp[e] $+$  gdp[p] $+$  gtp[e] $+$  gtp[p],
aso3[c] $+$  aso3[e] $+$  aso3[p] $+$  aso4[c], 
34dhpac[e] $+$  34dhpac[p] $+$  dopa[e] $+$  dopa[p],
tma[e] $+$  tma[p] $+$  tmao[e] $+$  tmao[p] \\ 
58  &  6  &  feoxam-un[c] $+$  feoxam-un[e] $+$  feoxam-un[p] $+$  feoxam[c] $+$  feoxam[e] $+$  feoxam[p] \\ 
59  &  6  &  cpgn-un[c] $+$  cpgn-un[e] $+$  cpgn-un[p] $+$  cpgn[c] $+$  cpgn[e] $+$  cpgn[p] \\ 
60  &  6  &  fecrm-un[c] $+$  fecrm-un[e] $+$  fecrm-un[p] $+$  fecrm[c] $+$  fecrm[e] $+$  fecrm[p]\\ 
61  &  6  &  fe3hox-un[c] $+$  fe3hox-un[e] $+$  fe3hox-un[p] $+$  fe3hox[c] $+$  fe3hox[e] $+$  fe3hox[p] \\ 
62  &  6  &  arbtn-fe3[c] $+$  arbtn-fe3[e] $+$  arbtn-fe3[p] $+$  arbtn[c] $+$  arbtn[e] $+$  arbtn[p] \\ 
63  &  6  &   acgal1p[e] $+$  acgal1p[p] $+$  acgal[e] $+$  acgal[p] $+$  udpacgal[e] $+$  udpacgal[p] \\ 
64  &  6  &  cu2[c] $+$  cu2[e] $+$  cu2[p] $+$  cu[c] $+$  cu[e] $+$  cu[p] \\ 
65  &  6  &  cyan[e] $+$  cyan[p] $+$  tcynt[e] $+$  tcynt[p] \\
66  &  6  &  chol[c] $+$  chol[e] $+$  chol[p] $+$  g3pc[c] $+$  g3pc[e] $+$  g3pc[p] \\ 
67  &  7  &  mercppyr[c] $+$  tcynt[c] $+$  tcynt[e] $+$  tcynt[p] $+$  tsul[c] $+$  tsul[e] $+$  tsul[p] \\ 
68  &  7  &  pac[c] $+$   pacald[c] $+$   pacald[e] $+$   pacald[p] $+$   peamn[e] $+$   peamn[p] $+$   phaccoa[c] \\ 
69  &  9  &  g3pi[c] $+$  g3pi[e] $+$  g3pi[p] $+$  inost[c] $+$  inost[e] $+$  inost[p] $+$  mi1p-D[c] $+$  minohp[e] $+$  minohp[p] \\ 
70  &  9  & 5prdmbz[c] $+$  adocbl[c] $+$  adocbl[e] $+$  adocbl[p] $+$  cbl1[c] $+$  cbl1[e] $+$  cbl1[p] $+$  dmbzid[c] $+$  rdmbzi[c] \\ 
71  &  10  & crnDcoa[c] $+$  crn-D[c] $+$  crn-D[p] $+$  crn[c] $+$  crn[e] $+$  crn[p] $+$  crncoa[c] $+$  ctbt[c] $+$  ctbt[p] $+$  ctbtcoa[c] \\ 
72  &  10  & (2) dopa[e] $+$  (2) dopa[p] $+$  (2) h2o2[e] $+$  (2) h2o2[p] $+$  o2s[e] $+$  o2s[p] $+$ (2) peamn[e] $+$  (2) peamn[p] $+$  (2) tym[e] $+$  (2) tym[p] \\ 
73  &  11  & aragund[c] $+$  garagund[c] $+$  gfgaragund[c] $+$  (2) o16a2und[p] $+$  (3) o16a3und[p] $+$  (4) o16a4colipa[e] $+$  (4) o16a4colipa[p] $+$  (4) o16a4und[p] $+$  o16aund[c] $+$  o16aund[p] $+$  ragund[c] \\ 
74  &  12  & adocbi[c] $+$   adocbip[c] $+$   adocbl[c] $+$   adocbl[e] $+$   adocbl[p] $+$   agdpcbi[c] $+$   cbi[c] $+$   cbi[e] $+$   cbi[p] $+$   cbl1[c] $+$   cbl1[e] $+$   cbl1[p]  \\ 
\hline
\end{tabular}
\caption{The 36 additional CMPs that are found in iAF1260 in a `minimal medium'.}
\label{pool2}
\end{table*}

We can argue that a suitable set of additional uptakes (comprising tRNAs, selenium, disulfide proteins, the aforementioned redox enzymes, biotin, and ACP) together with the missing periplasm-cytosol transport reactions (for cyan, tma and dms) will render the iAF1260 network completely open, thereby allowing for the possibility that the levels of each of the chemical moieties appearing are altered. Eliminating uptakes, on the other hand, will generically generate additional CMPs. Table \ref{pool2} reports the 36 extra CMPs that occur in iAF1260 in a `minimal medium' containing only ca2, fe2, glc-D, h2o, h, k, mg2, mn2, na1, nh4, o2, pi, so4 and zn2 (i.e. by allowing for 14 of the 299 possible uptakes). Detailed inspection reveals that many of these CMPs emerge from the lack of uptakes for elements like silver (39), cadmium (40), nickel (41), molybdenum (42), cobalt (43), tungsten (44), mercury (46), chloride(47), arsenic (55) and copper (64). A more detailed biochemical analysis is required to interpret the remaining pools. 

Notice that, while 72 of the CMPs discussed above correspond to solutions of (\ref{sys2}) with $k_m\in\{0,1\}$ $\forall m$, CMP 72 has $k_m\in\{0,1,2\}$ $\forall m$ while CMP 73 has $k_m\in\{0,1,\ldots,4\}$ $\forall m$. This shows that, while in general identifying CMPs cannot be treated as a Boolean problem, the range of values of $k_m$ to be considered in (\ref{sys2}) can be relatively small.

\subsection{iJR904}

For sakes of comparison, in Tables \ref{pooljr904} and \ref{pooljr904b} we report the CMPs found in the iJR904 metabolic network reconstruction of {\it E. coli} in the `rich' (all uptakes allowed) and `minimal' (defined in the same way as for iAF1260) media, respectively. One can see that, in essence, all of the CMPs of iJR904 are included among those of iAF1260, with some simplifications. For instance, the pool related to ACP conservation (17 in Table \ref{pooljr904}) is smaller in iJR904 than it is in iAF1260. Notice that CMP 29 in Table \ref{pooljr904b} displays two anomalous coefficients $k_m=50$, due to the non-integer stoichiometry with which the corresponding compounds occur in the network.

\begin{table*}
\begin{tabular}{  | c  | c  | p{13cm} | }  
\hline 
CMP ID & Size & Formula  \\ 
\hline 
1--10 & 2  & trdrd[c] $+$ trdox[c], 
seln[c] $+$ selnp[c],
trnaglu[c] $+$ glutrna[c], 
dms[c] $+$ dmso[c],
tmao[c] $+$ tma[c],
hqn[c] $+$ arbt6p[c],
tcynt[c] $+$ cyan[c],
3dhguln[c] $+$ 23doguln[c],
idp[c] $+$ itp[c],
acon\_T[c] $+$ aconm[c] \\
11--14 & 3 & ctbt[c] $+$ gbbtn[c] $+$ crn[c],
g3pi[c] $+$ inost[c] $+$ mi1p\_D[c],
8aonn[c] $+$ amob[c] $+$ pmcoa[c],
bbtcoa[c] $+$ crncoa[c] $+$ ctbtcoa[c]\\
15 & 4 & pacald[c] $+$ peamn[c] $+$ pac[c] $+$ phaccoa[c] \\
16 & 6 & pmcoa[c] $+$ 8aonn[c] $+$ dann[c] $+$ dtbt[c] $+$ btn[c] $+$ btnso[c] \\ 
17 & 12 & apoACP[c] $+$ acACP[c] $+$ actACP[c] $+$ ACP[c] $+$ malACP[c] $+$ ddcaACP[c] $+$ octeACP[c] $+$ myrsACP[c] $+$ palmACP[c] $+$ hdeACP[c] $+$ tdeACP[c] $+$ 3hmrsACP[c] \\ 
\hline
\end{tabular}
\caption{The 17 CMPs found for the network iJR904 in a Ôrich mediumÕ. The suffix [c] indicates that the compound occurs in the cytoplasm.\label{pooljr904}}
\end{table*}

\begin{table*}[ht!!!!!!]
\begin{tabular}{  | c  | c  | p{13cm} | }  
\hline 
Pool ID & Size & Formula  \\ 
\hline 
18--23 & 2 &  fuc1p\_L[c] $+$ fuc1p\_L[e],
dmso[e] $+$ dms[e],
nad[e] $+$ amp[e],
met\_D[e] $+$ met\_D[c],
tmao[e] $+$ tma[e],
gbbtn[e] $+$ crn[e] \\
24--27 & 3 &  glyb[c] $+$ betald[c] $+$ glyb[e],
taur[e] $+$ taur[c] $+$ aacald[c],
gbbtn[c] $+$ gbbtn[e] $+$ bbtcoa[c],
tsul[e] $+$ tsul[c] $+$ tcynt[c]\\
28 & 5 &  ctbtcoa[c] $+$  ctbt[c] $+$   crncoa[c] $+$  crn[c] $+$  crn[e]  \\
29 & 5 &  g3pc[c] $+$ chol[c] $+$  (50) pc\_EC[c] $+$  (50) agpc\_EC[c] $+$ chol[e]   \\
30 & 6 &  rdmbzi[c] $+$ adocbl[c] $+$ cbl1[c] $+$ cbl1[e] $+$ 5prdmbz[c] $+$ dmbzid[c]  \\ 
31 & 7 &  adocbip[c] $+$ agdpcbi[c] $+$ adocbl[c] $+$ cbl1[c] $+$ cbl1[e] $+$ adocbi[c] $+$  cbi[c]  \\
\hline
\end{tabular}
\caption{The 14 additional CMPs that are found in iJR904 in a Ôminimal mediumÕ. \label{pooljr904b}}
\end{table*}

Considering the `rich medium', it is interesting to note that, even though Table \ref{pooljr904} exhausts all of its CMPs, an additional conservation law emerges upon studying the left kernel of the stoichiometric matrix, whose dimension turns out to be 18 rather than 17.
This law violates (\ref{sysg}), i.e. it cannot be expressed through positive coefficients, and the corresponding conserved quantity is formed $7$ metabolites with the formula 5prdmbz[c] $+$ dmbzid[c] $+$ rdmbzi[c] $-$ adocbi[c] $-$ adocbip[c] $-$ agdpcbi[c] $-$ cbi[c]. Once we move to the `minimal medium', however, the metabolites pertaining to this conservation law fall into well defined CMPs, namely pools 30 and 31 in
Table \ref{pooljr904b}.  The additional 14 CMPs displayed by iJR904 in a `minimal medium' complete all the conservation laws of the network.

\subsection{Scaling of the number of conservation laws with the network size}

\begin{figure}[h!]
\begin{center}
\includegraphics*[width=.45\textwidth,angle=0]{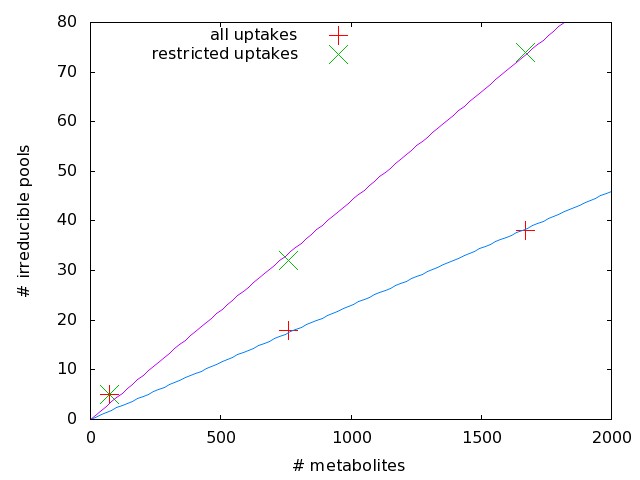}
\caption{Size of the pool basis as a function of $M$ for three {\it E. coli} metabolic network reconstructions (see text) in `rich' and `minimal' media.\label{figurapools}}
\end{center}
\end{figure}

In Figure \ref{figurapools} we show the size of the pool basis (i.e. the number of irreducible pools) as a function of the network size ($M$ in this plot) for the two networks we have considered so far and for a smaller {\it E. coli} network, namely the core matrix of the iAF1260 model (formed by $M=72$ metabolites that interact through $N=94$ reactions \cite{Feist}), both for the `rich' and `minimal' media\footnote{The `rich medium' for the {\it E. coli} core network allows for 20 uptakes, 5 of which (glc-D, o2, nh4, pi, h2o) survive in the `minimal medium'. The network is easily seen to present, in both media, 5 CMPs with straightforward biochemical meaning: nad $+$ nadh (nad conservation), nadp $+$ nadph (nadp conservation), q8 $+$ q8h2 (coenzyme-Q conservation), amp $+$ arp $+$ atp (adenylate moiety conservation) and coa $+$ accoa $+$ succoa (coenzyme-A conservation).}.  It appears that the number of irreducible pools scales approximately linearly with the network size.

While the investigation of a larger family of networks is needed to characterize this regularity more thoroughly, some insight can already be obtained from the analysis of random networks. We consider, in particular, an ensemble of `random metabolic networks' with $N$ reactions and $M$ compounds, such that each stoichiometric coefficient is chosen randomly and independently with probabilities ($0<\gamma<1$)
\begin{gather}
\label{ensemble}
\text{Prob}(S_{mr}=0) = 1-\gamma \\
\text{Prob}(S_{mr}=1)=\text{Prob}(S_{mr}=-1)=\gamma/2~~.
\end{gather}
The parameter $\gamma$ rules the average connectivity of the network. In particular, we will assume that $\gamma = 2c/N$ (with $c$ a constant), and consider the limit $N,M \to \infty$, keeping a constant ratio $\alpha=N/M$. In this limit, the above model generates a Poisson distribution for the in- and out-degrees, with average values $\langle k_{{\rm in}}^m \rangle = \langle k_{{\rm out}}^m \rangle = c$ for metabolites and $\langle k_{{\rm in}}^r \rangle = \langle k_{{\rm out}}^r \rangle = c/\alpha$ for reactions, respectively. (In real networks, the degree distribution of metabolites  is known to have heavy tails due to the presence of ubiquitous compounds like water, ATP, etc. whose connectivity typically grow with the network size. On the other hand, the degree distribution of the remaining metabolites follows a Poissonian to a good approximation.) By definition, the overall number of pools of size $n$ is given by
\begin{equation}
\label{poolsL}
\mathcal{N}_n = \sum_{\mathbf{k} : ||\mathbf{k}||=n}~ \prod_{r=1}^N \delta\left(\sum_{m=1}^M k_m S_{mr}\right)~~,
\end{equation}
where $\delta(x)$ denotes Dirac's $\delta$-function, we assume $k_m\in\{0,1\}$ for sakes of simplicity, and $||\mathbf{k}||=\sum_m k_m$. $\mathcal{N}_n$ as defined above however includes all linear combinations of irreducible pools that produce a pool of size $n$. To
obtain the number of irreducible pools, one should subtract from $\mathcal{N}_n$ the contributions due to superpositions of smaller pools. For instance, all distinct pairs of pools of size 1 would contribute to $\mathcal{N}_2$ as well, so that the number of irreducible pools of size $2$ is given by 
\begin{equation}\label{corre}
\mathcal{N}^{{\rm irr}}_2 = \mathcal{N}_2 - \frac{\mathcal{N}_1 (\mathcal{N}_1-1)}{2}~~. 
\end{equation}
Expression (\ref{poolsL}) furthermore depends on the particular network being examined. We shall focus on its average over the entire ensemble. Writing the $\delta$-function as $\delta(x) = \frac{1}{2\pi}\int_{0}^{2 \pi} \exp({\rm i} x \phi)d\phi$, summing over $k_m$'s and averaging over the stoichiometry one finds  
\begin{equation}
\langle \mathcal{N}_n \rangle = \binom{M}{n}\left[\frac{1}{2\pi} \int_{0}^{2\pi}  (1-\gamma+\gamma\cos\phi)^n d\phi  \right]^N~~ .
\end{equation} 
Expanding the integrand and noting that 
\begin{equation}
\frac{1}{2\pi}\int_{0}^{2 \pi} (\cos \phi)^k d\phi  = 
\begin{cases}
2^{-k}\binom{k}{k/2} &\text{if $k$ is even}\\
0 &\text{otherwise} 
\end{cases}~~,
\end{equation}
one finally obtains
\begin{equation}
\label{pulz}
\langle \mathcal{N}_n \rangle = \binom{M}{n}(1-\gamma)^{Nn}\left[\sum_{k~\textrm{even}}  \binom{n}{k}\binom{k}{k/2} \frac{\gamma^k}{2^k (1-\gamma)^k}\right]^N~~,
\end{equation}
which can be evaluated in the limit $N\to\infty$. For $n=2$, keeping only the leading-order terms in $M$ and approximating $\langle \mathcal{N}_1^2\rangle \simeq \langle \mathcal{N}_1 \rangle^2$ in (\ref{corre}), one gets
\begin{equation}
\langle\mathcal{N}^{{\rm irr}}_2\rangle\simeq \frac{M}{2}(e^{-2c}-e^{-4c})~~.
\end{equation}
For the networks being examined, once the most connected compounds are removed, one finds $c \simeq 1.6$, leading to $\langle\mathcal{N}^{{\rm irr}}_2\rangle \simeq 2, 15, 32$ for the {\it E. coli} core, iJR904 and iAF1260 models, respectively (considering a `rich medium'). This should be compared with the actual numbers of irreducible pools of size 2 we found, namely $3, 10$ and $31$ respectively. (Similar results can be obtained, with more work, for larger values of $n$.)

It is now straightforward to show that the size $B$ of the pool basis scales linearly with $M$. Upon summing (\ref{pulz}) over $n$, the total number of pools $\mathcal{N}_{{\rm tot}}=\sum_n \langle \mathcal{N}_n \rangle$ is seen to satisfy
\begin{equation}
(1+e^{-2c})^M \leq \mathcal{N}_{{\rm tot}} \leq 2^M~~.
\end{equation}
In other terms, there exists a number $z \in [1,2]$ such that $\mathcal{N}_{{\rm tot}} = z^M$. On the other hand, assuming for simplicity that pools in the basis are non-overlapping, one has $\mathcal{N}_{{\rm tot}} = 2^B$, from which we get $B  = M \log_2 z$, i.e. a linear scaling with $M$, in agreement with the behavior displayed in Figure \ref{figurapools}.

\section{Materials and Methods} 
\label{sec:matmet}

As outlined above, our strategy employs a message-passing method (namely BP) to identify a list of metabolites belonging to at least one irreducible pool, a Monte Carlo method to extract individual irreducible pools from the list, and a relaxation algorithm to check that all irreducible pools have been obtained. We shall now discuss each procedure in more detail.

\subsection{Monte Carlo}

The problem of finding the integer solutions of (\ref{sys2}) can be mapped onto that of finding the ground states of a fictitious, discrete `energy function' given by
\begin{equation}\label{statmech}
E (\mathbf{k}) =\sum_{m,n=1}^M J_{m,n} k_m k_n \geq 0~~,
\end{equation}
where
\begin{equation}\label{coupling}
J_{m,n}\equiv \sum_{i=1}^N S_{m,i} S_{n,i} ~~ .
\end{equation}
Evidently, $E(\mathbf{k})=0$ if $\mathbf{k}$ satisfies (\ref{sys2}), so that CMPs correspond to vectors $\mathbf{k}$ for which $E=0$. Several optimized Monte Carlo methods (like simulated annealing) are available to compute the minima of functions like (\ref{statmech}) \cite{montecarlokrauth}. More specifically, these methods are capable of sampling vectors $\mathbf{k}$ distributed according to
\begin{equation}\label{boltz}
P(\mathbf{k}) = \frac{e^{-E(\mathbf{k})/T}}{Z(T)}~~,
\end{equation} 
where $T>0$ is a parameter (the fictitious `temperature' of the system) and $Z(T)$ is a normalization factor. The simplest controlled method to generate configurations according to (\ref{boltz}) is probably the Metropolis scheme: at each step, select a variable $k_m$ at random among those appearing in the list of metabolites belonging to at least one CMP obtained by BP (see below), and propose an update of the form $k_m \to k_m + \delta$, where $\delta=\pm 1$ with equal probability if $k_m>0$ and $\delta=1$ if $k_m=0$. Next, evaluate the ensuing change of $E$, i.e.
\begin{equation}
\Delta E = 2 \delta \sum_{n \neq m} J_{m,n} k_n + J_{m,m}~~.
\end{equation}
The proposed move is then accepted with probability
\begin{equation}
P({\rm accept}) = \textrm{arg }\min~\{1,e^{-\Delta E/T}\}~~.
\end{equation}
For each choice of $T$, this Markov chain converges to (\ref{boltz}) with a mixing time that increases as $T$ gets smaller and depends on the distribution of initial states (so that the closer the latter is to (\ref{boltz}), the smaller the mixing time) \cite{montecarlobook}. The minima of $E$ are recovered in the limit $T\to 0$, which can be achieved operationally by initializing the Monte Carlo simulation at some large value of $T$ and then decreasing the `temperature' at a constant rate until $T=0$ (note that for $T=0$ the above dynamics becomes a gradient descent). When lowering $T$, it is possible to speed up the convergence by using, as the initial state for a certain $T$, the final state at the previous (higher) value of $T$.

In this work, we have relied on Monte Carlo in order to find integer solutions of (\ref{sys1}). In specific, in order to retrieve the conserved pools as ground states, we have performed iterated Metropolis-based annealings to minimize the energy (\ref{statmech}). While Monte Carlo is generically a costly procedure (with running times growing as a high power of the number of variables), in our case CPU costs are limited by the fact that the search space includes a number of metabolites much smaller than $M$.

\subsection{Belief Propagation}

Message passing algorithms are efficient computational strategies which are exact on tree graphical models \cite{pearl1982,kim1983}, and are extensively used as a heuristic procedure to solve problems defined on sparse graphs \cite{Yedidia,kschischang2001, BMZ}. Here they provide a convenient technical mean to obtain a refined list of metabolites belonging to at least one CMP. To present their implementation in the present case (BP), let us introduce the cost function (see again (\ref{sys2}))
\begin{equation}
\label{eq:costfun}
H(\mathbf{k}) = \sum_{i=1}^N \left[1-\delta\left( \sum_{m=1}^M S_{m,i} k_m ; 0\right)\right]\,\,\,\,,
\end{equation}
where $\delta(x;y)$ is the Kronecker delta function ($=1$ if $x=y$ and $= 0$ otherwise). One easily understands that all vectors $\mathbf{k}$ such that $H=0$ are also zero-energy configurations for (\ref{statmech}), and vice-versa.  In essence, BP aims at computing the marginals of the probability distribution
\begin{equation}
\label{eq:probk}
P({\mathbf k})  = \lim_{T \rightarrow 0} \frac{e^{-H(\mathbf{k})/T}}{Z(T)} = \frac {1}{\mathcal{N}_{\mathrm{sol}}} \prod_{i=1}^N \delta \left(\sum_{m=1}^M S_{m,i} k_m ; 0\right) ~~,
\end{equation}
(where ${\cal N}_\mathrm{sol}$ stands for the number of solutions of (\ref{sys2})), i.e. the probability $P_m(k_m)$ that the $m$-th coordinate of the $\mathbf{k}$ vector attains a value $k_m$ over the solution space of (\ref{sys2}). It is easy to understand that disposing of such marginals is equivalent to disposing of the list of metabolites belonging to at least one CMP (actually, the marginals encode for more information). Note that, by definition,
\begin{equation}
\label{eq:marginal1}
P_m(k_m) \equiv \sum_{\{k_n \}_{n \neq m}}
P({\mathbf k}) = \frac1{{\cal N}_{\mathrm{sol}}} \sum_{s=1}^{{\cal N}_{\mathrm{sol}}} \delta(k_m^s; k_m) ~~,
\end{equation}
where in the last equivalence we stressed the probabilistic interpretation of the marginal as the histogram of the $m$-th coordinate over the solutions $\{{\bf k}^s\}$, $s =1, \dots,{\cal N}_{\mathrm{sol}}$. A direct evaluation of (\ref{eq:marginal1}), however, would require computing a sum over $k_{{\rm max}}^{M-1}$ terms (assuming that $k_m\in\{0,1,2,\ldots, k_{{\rm max}}\}$), which becomes computationally unfeasible for $M$ larger than a few tens. BP allows to overcome such a severe computational load at the cost of introducing an approximation that makes the algorithm formally exact only on locally tree-like networks \cite{montanari-mezard-book} (it is however known that it can be applied also on loopy graphs, like metabolic networks). The algorithm is based on nodes (metabolites and reactions) exchanging two types of `messages':
\begin{itemize}
\item $\mu_{i \rightarrow m}(k)$ (from reactions to metabolites): the (non-normalized) probability that the constraint imposed by the $i$-th equation in (\ref{sys2}), i.e. $\sum_n S_{n ,i}k_n = 0$, is fulfilled given that the $m$-th variable takes value $k$;
\item $\rho_{m \rightarrow i} (k)$ (from metabolites to reactions): the probability that the $m$-th variable takes value $k$ in the absence of reaction $i$.
\end{itemize}
One easily sees that these quantities are related by
\begin{gather}\label{eq:bp1}
\mu_{i \rightarrow m}(k_m) = \sum_{\{k_n \}_{n \neq m}}\delta\left(\sum_{n=1}^M S_{n ,i}k_n ; 0\right) \prod_{n \in i
  \setminus m} \rho_{n \rightarrow i} (k_n ) \\
\rho_{m \rightarrow i} (k_m) = C_{m \rightarrow i}\prod_{ j
  \in m \setminus  i} m_{j \rightarrow m}(k_m) 
\label{eq:bp}
\end{gather}
where $C_{m \rightarrow i}$ is a constant enforcing the normalization of the probability $\rho_{m \rightarrow i} (k_m)$, 
and the subscript $j \in m \setminus i$ denotes the set of reactions producing or consuming metabolite $m$ but reaction $i$ (and similarly for the subscript $n\in i \setminus m$). The above equations can be solved iteratively by initializing messages at random and updating them in a random sequential order. The algorithm halts when the difference between each message at iteration $t$ and iteration $t-1$ is less than a pre-defined threshold ($10^{-8}$ in all our simulations). While the number of iterations is generically problem-dependent, in the simulations presented here it rarely exceeds a few hundreds. Once we have reached convergence in the update, the desired marginal probability distribution is computed as the product of {\em all} $m$-messages pointing to the $m$-th variable, namely
\begin{equation}
\label{eq:marginal2}
P_m(k_m) = C_m \prod_{j \in m} \mu_{j \rightarrow m}(k_m)~~,
\end{equation}
where $C_m$ is, again, a normalization constant that is easily computed. One can see that, formally, (\ref{eq:bp1}), (\ref{eq:bp}) and (\ref{eq:marginal2}) only hold when the underlying network is a tree \cite{montanari-mezard-book}. However, the same equations can be employed to study problems like (\ref{sys2}) on more complicated networks, the key correctness test being their ability to find solutions.

By using BP, it is possible to compute conditional marginal probability distributions, i.e. the marginal probability distribution that the $m$-th variable takes value $k$ given that a set of $\ell$ variables $\{n _1,\dots, n _\ell\}$
take values $\hat k_{n _1},\dots, \hat k_{n _\ell}$ respectively, by simply adding `external fields' to the cost function (\ref{eq:costfun}), e.g.
\begin{equation}
\label{eq:costfunfields}
H({\bf k}) \to H({\bf k}) + T\sum_{n =1}^M h_n  \delta(k_{n }; \hat k_n )~~,
\end{equation}
where the weight $T$ has been included so as to ensure a well behaved $T \rightarrow 0$ limit in (\ref{eq:probk}). By setting
\begin{equation}
h_n=
\begin{cases}
h_0\gg 1 &\text{for $n \in \{n _1,\dots, n _\ell\}$}\\
0 &\text{otherwise}
\end{cases}
\end{equation}
we are able to set variables $k_{n _1},\dots, k_{n _n}$ to our desired values $\hat k_{n _1},\dots, \hat k_{n _\ell}$ and thus force the BP iteration to converge to the correct conditional probability distribution.

Upon convergence, all marginal probability distributions are evaluated using (\ref{eq:marginal2}) and variables are ranked according to their polarization (i.e., in terms of how peaked the probability is around its most probable value). The strongly polarized variables are then forced to take the most probable value (different from zero) as  explained above, and added to the list of metabolites that putatively belong to at least one pool. Once the first polarized variable is fixed, the algorithm is iterated until no other variables are polarized, i.e. until all the marginal probability distributions of the remaining `free' variables are concentrated at zero (i.e. none of them belongs to a CMP). 
The final result is thus a list of metabolites belonging to at least one CMP. To account for dependencies on the initialization of messages, we have repeated the procedure for 20 different choices of the initial conditions, and used, as the final list, the union of the different outputs. Note however that the lists obtained by different initial conditions varied at most by few metabolites (less than 4) among each other.

\subsection{Relaxation algorithm}


To check that all metabolites belonging to at least one CMP have been found, we remove the corresponding rows from the stoichiometric matrix $\mathbb{S}$ and look for a solution of (\ref{sys3}) with strict inequalities (exploiting Motzkin's theorem). To this aim, we resort to relaxation algorithms \cite{shrij}. These classic methods work by correcting iteratively ($t$ being the step) the least unsatisfied constraint, according to the scheme
\begin{eqnarray}
m_t ={\rm arg}~\min_m \sum_i S_{m,i} v _i(t) \label{luc} \\
v _i(t+1) = v _i(t) +q S_{i,m_t}~~,
\end{eqnarray}  
$q$ being a parameter that can be fixed in different ways, from a constant (as in MinOver \cite{minover0}) to a quantity proportional to the amount by which the constraint is violated (as in the so-called Motzkin scheme \cite{shrij}). The above dynamics converges to a solution, if one exists, in polynomial time. If relaxation doesn't converge, the reduced matrix $\mathbb{S}$ contains CMPs that BP was unable to identify (we recall that BP is approximate for loopy networks). In this case, the missing metabolites can be found by analyzing the relaxation dynamics in detail. As discussed elsewhere \cite{noiplos, deMartino12}, after a transient, the algorithm visits frequently the constraints that prevent convergence. Therefore, they are easily identified by keeping tracks of the least unsatisfied constraints (\ref{luc}) over time.

\section{Discussion}

Conservation laws (described by the left kernel of the stoichiometric matrix $\mathbb{S}$) take on a specific biochemical significance when the coefficients involved are non-negative integers, in which case each relation describes the conservation of a particular molecular moiety. In turn, identifying the irreducible conserved moieties embedded in a given $\mathbb{S}$ requires solving the hard constraint-satisfaction problem of finding all integer, non-negative solutions to a linear system of equations defined from the network's input-output relationships. Methods allowing to solve this problem on small networks are relatively straightforward and have been studied before. For the network sizes relevant in metabolic modeling, however, the {\it a priori} search space of the problem is huge and exact methods are doomed to fail due to exceeding computational costs. Luckily, remarkably powerful heuristics to tackle this type of problems has been developed in the last decade at the interface between statistical mechanics and computer science. Here, starting from one such procedure, we have constructed and applied a technique that allows to obtain full information about the irreducible CMPs for a given $\mathbb{S}$. Our method, in particular, combines BP (used to reduce the size of the search space) with Monte Carlo (used to disentangle individual pools) and a relaxation method (used to test that all irreducible pools have been found) and allows to retrieve a detailed map of conserved moieties in genome-scale networks. We have analyzed the structure of the CMPs emerging in two large-scale reconstructions of the metabolism of {\it E. coli}. In most cases, CMPs either display a simple biochemical meaning or their origin can be clearly traced back to properties of the reconstruction. In other cases, however, it is difficult to identify a precise rationale for the groups we obtain. More generally, we have suggested the existence of a linear relation between the number of irreducible CMPs and the network size, validating it in data and by an analytical calculation for `random metabolic networks' (although more work will be needed to characterize this picture more thoroughly).

Besides their importance for dynamical modeling widely discussed in
the literature \cite{Bakker, Rao, Sauro, Famili}, CMPs provide crucial
indications concerning how a cell will respond to a perturbation that
e.g. increases the level of a particular chemical species. The manner
in which that perturbation propagates is indeed constrained by the map
of CMPs. In addition, results obtained by the method we propose can
improve producibility predictions \cite{Carlotta}. The technique we have
presented is successful in large, genome-scale models, so that applicability to
other organisms is straightforward. More interestingly, however, it
could represent a general protocol by which different stoichiometry-based
problems that are inherently integer programming ones can be tackled.
On the negative side, because of the many different steps involved, it
is hard to envision an automatization at this point in time. It is
therefore important to explore alternatives that, for instance, might
eliminate the Monte Carlo step. A viable possibility could be
to complement BP with a decimation procedure \cite{ZecchinaScience,
  KrzakalaPNAS2007}.


~


The authors thank A. Braunstein for sharing with us the implementation of the BP algorithm, and for many relevant discussions. RM acknowledges financial support from the Alexander von Humboldt foundation. This work is supported by the DREAM Seed Project of the Italian Institute of Technology (IIT). The IIT Platform Computation is gratefully acknowledged.

 

\bibliographystyle{unsrt}   
\bibliography{refpools-2}      

\end{document}